\documentclass[preprint,preprintnumbers,nofootinbib]{revtex4}

\usepackage[dvips]{color}
\usepackage{graphicx}
\usepackage{bbm}

\newcommand{\hif}{\mathchar`-}

\newcommand{\ev}{{\rm eV}}
\newcommand{\kev}{{\rm keV}}

\newcommand{\gev}{{\rm GeV}}

\newcommand{\bmx}{\left(\begin{array}}
\newcommand{\emx}{\end{array}\right)}

\newcommand{\true}{{\bf 1}}
\newcommand{\ip}{{\bf 1}^{'}}
\newcommand{\ipp}{{\bf 1}^{''}}
\newcommand{\ippp}{{\bf 1}^{'''}}
\newcommand{\two}{{\bf 2}}
\newcommand{\tp}{{\bf 2}^{'}}

\begin{document}


\title{$Q_6$ flavor symmetry model for the extension of the minimal standard
model by three right-handed sterile neutrinos}

\author{
Takeshi Araki\footnote{araki@ihep.ac.cn} and Y.F.
Li\footnote{liyufeng@ihep.ac.cn} } \affiliation{ Institute of High
Energy Physics, Chinese Academy of Sciences, Beijing 100049, China }

\begin{abstract}
The extension of the minimal standard model by three right-handed
sterile neutrinos with masses smaller than the electroweak scale
($\nu$MSM) is discussed in a $Q_6$ flavor symmetry framework. The
lightness of the $\kev$ sterile neutrino and the near mass
degeneracy of two heavier sterile neutrinos are naturally explained
by exploiting group properties of $Q_6$. A normal hierarchical mass
spectrum and an approximately $\mu$-$\tau$ symmetric mass matrix are
predicted for three active neutrinos. Nonzero $\theta_{13}$ can be
obtained together with a deviation of $\theta_{23}$ from the
maximality, where both mixing angles are consistent with the latest
global data including T2K and MINOS results. Furthermore, the tiny
active-sterile mixing is related to the mass ratio between the
lightest active and lightest sterile neutrinos.

\end{abstract}

\maketitle

\section{introduction}
Both the establishment of neutrino oscillation phenomena and the
evidence of nonluminous dark matter (DM) demand physics beyond the
standard model (SM). On the one hand, compelling evidences from
current solar, atmospheric, reactor and accelerator neutrino
experiments have told us that neutrinos are massive and lepton
flavors are mixed \cite{PDG}. On the other hand, various
cosmological observations have revealed that DM is five times more
abundant than normal matter and accounts for about one quarter of
the Universe (i.e., $\Omega_{\rm DM}\simeq0.227\pm0.014$
\cite{WMAP10}). To explain both problems above, an extension of the
SM by three right-handed sterile neutrinos with masses smaller than
the electroweak scale ($\nu$MSM) was first proposed by Asaka {\it et
al.} \cite{NuMSM1}. The smallness of active neutrino masses is
described by the canonical seesaw mechanism \cite{type1} and one
light right-handed sterile neutrino at the $\rm keV$ scale acts as a
candidate of warm dark matter (WDM). Moreover, the model can explain
the baryon asymmetry in the Universe through the oscillations
\cite{BANO,BAU} of two heavier right-handed sterile neutrinos with
masses at the $\gev$ scale.

Despite the above phenomenological successes, one unsatisfactory
point in the $\nu$MSM is the lack of a natural explanation for large
mass splitting between the $\kev$ sterile neutrino and the heavier
ones. Moreover, the oscillation mechanism for the baryon asymmetry
demands strong mass degeneracy between the heavier sterile
neutrinos, but the $\nu$MSM is impotent to be able to explain its
origin. In order to resolve theses issues, some interesting ideas
have been proposed in the context of a $\rm U(1)$ flavor symmetry
\cite{U1FS}, the Froggatt-Nielsen mechanism \cite{FN}, the split
seesaw mechanism \cite{Split} and grand unified theories \cite{CR}.
In the present work, we introduce a non-Abelian discrete flavor
symmetry and try to understand the mass splitting and degeneracy due
to group properties of the flavor symmetry. The mass degeneracy of
two heaver sterile neutrinos could be interpreted as a sign that
they constitute a doublet representation of the flavor symmetry,
while it may be natural to assign a singlet representation to the
$\kev$ sterile neutrino. Furthermore, if the singlet representation
is a complex one, we can prohibit a bare mass term of the $\kev$
sterile neutrino because of the Majorana nature and may be able to
generate a suppressed mass term from higher-dimensional operators.
Inspired by these clues, we employ the $Q_6$ group as our flavor
symmetry\footnote{Previous studies about $Q_6$ can be found in Ref.
\cite{kubo}.} since it is the smallest finite group\footnote{See
Ref. \cite{Frampton} for the classification of discrete groups upto
$g=31$, and Ref. \cite{FSrev} for the recent reviews of non-Abelian
discrete flavor symmetries.} which contains both a complex singlet
and a real doublet representations. In addition to $Q_6$, we
introduce two auxiliary $Z_{N}$ symmetries in order to handle the
order of magnitude of some small parameters as well as to forbid
unwanted terms. We find that our model predicts a normal
hierarchical mass spectrum and an approximately $\mu$-$\tau$
symmetric mass matrix for three active neutrinos.

The remaining part of this paper is organized as follows. In section
II we present the framework of the $\nu$MSM with a $Q_6 \times Z_3
\times Z_2$ flavor symmetry. Section III is devoted to the
realization of the seesaw mechanism and section IV to the
diagonalization and the resulting neutrino masses and mixing matrix.
A numerical analysis with the focus on nonzero $\theta_{13}$ is also
illustrated in section IV. Finally, we give a summary in
section V.

\section{The $Q_6 \times Z_3 \times Z_2$ Model}
\begin{table}
\begin{tabular}{|c|c|c|c|c|c|c|c|c|c|c|c|}\hline
      & $L_1$   & $L_D$   & $E_1$   & $E_D$  & $N_1$  &
$N_D$ & $H$     & $S_x$  & $S_y$ & $S_z$ & $D$ \\ \hline
$Q_6$ & $\true$ & $\tp$   & $\ip$   & $\two$ & $\ipp$ &
$\tp$ & $\true$ & $\ipp$  & $\ippp$ & $\true$ & $\tp$ \\ \hline
$Z_3$ & $0$     & $0$     & $2$     & $1$    & $0$    &
$0$   & $0$     & $2$     & $2$     & $0$ & $0$ \\ \hline
$Z_{2}$ & $0$     & $0$     & $0$     & $0$    & $0$    &
$1$ & $0$     & $0$     & $0$     & $1$ & $1$ \\ \hline
\end{tabular}
\caption{Particle content and charge assignments.}
\end{table}
We introduce three right-handed sterile neutrinos $N_{1,2,3}$ and
gauge-singlet flavon fields $S_x$, $S_y$, $S_z$ and $D$ with a $Q_6
\times Z_3 \times Z_{2}$ flavor symmetry. We assign a $Q_6$-singlet
(-doublet) for the first (second and third) generation of fermions,
and the SM Higgs is assumed to be invariant under all the flavor
symmetries. The particle content and charge assignments for each
symmetry are summarized in Table I, and the basic group theory of
$Q_6$ is reviewed in the appendix. Because of the symmetries, there
are no renormalizable Yukawa interactions in the charged lepton
sector, and the charged lepton masses follow from the higher
dimensional operators:
\begin{eqnarray}
{\cal L}_{\ell}=
 \frac{Y_x}{\Lambda}
 ( \overline{L}_D^{} H E_D^{} )_{\ippp}^{} S_x^{}
+\frac{Y_y}{\Lambda}
 ( \overline{L}_D^{} H E_D^{} )_{\ipp}^{} S_y^{}
+\frac{Y_e}{\Lambda^2_{}}~\overline{L}_1^{}
 H E_1^{} S_x^2
+ h.c.~,
\label{eq:Ll}
\end{eqnarray}
where we have specified the Lagrangian upto the next-to-leading
order level, and the charged lepton mass matrix is diagonal upto
this order. The subscripts beside $(\cdots)$ indicate the required
$Q_6$ representations to make the terms invariant under $Q_6$.
In the neutrino sector, $N_1$ is assumed to be a complex representation of $Q_6$, while the other leptons are real representations.
Consequently, only $S_x$ and $S_y$, which are complex representations, can reproduce a Dirac and Majorana mass terms for $N_1$, and those interactions are suppressed due to the $Z_N$ symmetries:
\begin{eqnarray}
&&{\cal L}_\nu =
 \frac{\alpha}{\Lambda^{}}~
 \overline{L}_1^{} {\tilde H} (N_D^{} D^{})_{\true}^{}
+\frac{\beta}{\Lambda^{}}
 (\overline{L}_D^{} {\tilde H} N_D^{})_{\tp}^{} D
+\frac{\gamma}{\Lambda^{}}
 (\overline{L}_D^{} {\tilde H} N_D^{})_{\true}^{} S_z^{}
\nonumber \\
&&\hspace{2cm}
+\frac{\delta}{\Lambda^3}~
 \overline{L}_1^{} {\tilde H} N_1^{} S_x^3
+\frac{\epsilon}{\Lambda^3}~
 (\overline{L}_D^{} {\tilde H} N_D^{})_{\ip}^{} S_x^{} S_y^{*} S^{}_z
+ h.c.~, \label{eq:Ld} \\
&&{\cal L}_M =
 m_a(N_D^{} N_D^{})_{\true}^{}
+\frac{m_b}{\Lambda^2}
 (N_D^{} N_D^{})_{\tp}^{} (D^2)_{\tp}^{}
+\frac{m_c}{\Lambda^2}~
  N_1^{} N_1^{} S_x^{} S_y^*
+h.c.~.
\label{eq:Lm}
\end{eqnarray}
Note that in the above expressions we have omitted some terms whose
contributions can be embedded into others and implicitly assumed a
mechanism which supplies $m_{a,b,c} = {\cal O}(1)~\gev$, e.g.,
the spontaneous breaking of a lepton number $Z_N$ symmetry at a $\gev$ scale.

We define the vacuum expectation values(VEVs) of neutral scalars as
\begin{eqnarray}
\langle H^0_{} \rangle = v = 174~\gev,~~
\langle S_x \rangle = s_x,~~
\langle S_y \rangle = s_y,~~
\langle S_z \rangle = s_z,~~
\langle D \rangle = (d_1,d_2), \label{eq:vev}
\end{eqnarray}
leading to the charged lepton masses,
\begin{eqnarray}
m_e^{}=
\left(\frac{s_x}{\Lambda}\right)^2 Y_e v,~~
m_\mu^{}=\frac{1}{\Lambda}
\left( Y_x s_x + Y_y s_y \right)v,~~
m_\tau^{}=\frac{1}{\Lambda}
\left( Y_x s_x - Y_y s_y \right) v~,
\label{eq:me}
\end{eqnarray}
the right-handed Majorana neutrino mass matrix,
\begin{eqnarray}
M_R=
\bmx{ccc}
 0 & 0 & 0 \\
 0 & 0 & m_a \\
 0 & m_a & 0
\emx
+
\frac{1}{\Lambda^2}
\bmx{ccc}
 m_c s_x s_y & 0 & 0 \\
 0 & m_b d_2^2 & 0 \\
 0 & 0 & m_b d_1^2
\emx ,
\end{eqnarray}
and the Dirac mass matrix,
\begin{eqnarray}
M_D
=
\frac{1}{\Lambda}
\bmx{ccc}
 0 & \alpha d_2 & \alpha d_1 \\
 0 & \beta d_1  & \gamma s_z \\
 0 & \gamma s_z & \beta d_2
\emx v
+
\frac{1}{\Lambda^3}
\bmx{ccc}
 \delta s_x^3 & 0 & 0 \\
 0 & 0 & \epsilon s_x s_y s_z \\
 0 & -\epsilon s_x s_y s_z & 0
\emx v.
\end{eqnarray}
Notice that, in Eq. (\ref{eq:me}), we have the same fine-tuning
problem as that in \cite{GriLav} for obtaining $m_\mu \gg m_\tau$.
Nevertheless, we will not tackle this problem and only focus on the
neutrino sector in what follows.
We also note that although the second terms of $M_R$ and $M_D$ are
strongly suppressed by $1/\Lambda^2$ compared with the first terms,
we keep $m_c$, $m_b$ and $\delta$ in our discussions because
they will be important when we discuss the sterile neutrino masses
and active-sterile mixing. In contrast, $\epsilon$ in $M_D$ only
contributes to the masses and mixing of three active neutrinos, and
its effects are negligibly small in comparison with the first term.
Thus, we shall ignore $\epsilon$.

\section{SeeSaw Mechanism}
Let us move on to the diagonal basis of $M_R$. The diagonalization
can approximately be done by the $45^\circ$ rotation in the $2\hif
3$ plane, and three sterile neutrino masses are found to be
\begin{eqnarray}
M_1 \simeq \frac{s_x s_y}{\Lambda^2} m_c,~~
M_2 \simeq m_a - \frac{m_b}{2\Lambda^2}(d_1^2 + d_2^2),~~
M_3 \simeq m_a + \frac{m_b}{2\Lambda^2}(d_1^2 + d_2^2)~.
\end{eqnarray}
$M_1$ is suppressed with $1/\Lambda^2$, and thus it is the candidate
of WDM, while $M_2$ and $M_3$ are nearly degenerate. Interestingly,
the order of $M_1$ and that of the mass difference between $M_2$ and
$M_3$ are the same, which may turn out to be a key ingredient when
one considers the baryon asymmetry in the Universe \cite{BAU}.
Nevertheless, we shall naively assume $M_2=M_3=m_a$ and do not
consider the baryon asymmetry in what follows since detailed studies
of the baryon asymmetry go beyond the scope of this paper. Because
of $M_1\ll M_{2,3}$, the masses of the light neutrinos are obtained
by integrating out only $M_2$ and $M_3$, yielding the following $4
\times 4$ effective mass matrix:
\begin{eqnarray}
M_{\nu}^{4\times 4} =
\bmx{cc} -{M}^{3\times 3}_{\nu} & \Delta \\
\Delta^{T} & M_1 \emx, \label{eq:4by4}
\end{eqnarray}
where
\begin{eqnarray}
M_\nu^{3\times 3}
&=&\frac{v^2}{m_{a}} \bmx{ccc} 2\alpha^{\prime 2} &
\alpha^{\prime}(\beta^{\prime}+\gamma^{\prime}) &
\alpha^{\prime}(\beta^{\prime}+\gamma^{\prime}) \\
\alpha^{\prime}(\beta^{\prime}+\gamma^{\prime}) &
2\beta^{\prime}\gamma^{\prime} &
\beta^{\prime 2}+\gamma^{\prime 2} \\
\alpha^{\prime}(\beta^{\prime}+\gamma^{\prime}) & \beta^{\prime
2}+\gamma^{\prime 2} & 2\beta^{\prime}\gamma^{\prime}
\emx\nonumber \\
&+& \varepsilon_{d} \frac{v^2}{m_{a}} \bmx{ccc} 0 &
\alpha^{\prime}(2\beta^{\prime}-\gamma^{\prime}) &
-\alpha^{\prime}(2\beta^{\prime}-\gamma^{\prime}) \\
\alpha^{\prime}(2\beta^{\prime}-\gamma^{\prime}) &
2\beta^{\prime}\gamma^{\prime} & 0 \\
-\alpha^{\prime}(2\beta^{\prime}-\gamma^{\prime}) & 0 &
-2\beta^{\prime}\gamma^{\prime} \emx + {\cal O}(\varepsilon^{2}_{d})
\label{eq:3by3}
\end{eqnarray}
will later end up the mass matrix of three active neutrinos, and
$\Delta=(\delta^{\prime} v,~0,~0)^T$ stands for the mixing effect of
active-sterile neutrinos. In the above expressions, we have defined
$d_{1}+d_{2}=2d$ and $d_{1}-d_{2}=2\varepsilon_{d}d$
and embedded the suppression factor $1/\Lambda$ into the couplings such that
\begin{eqnarray}
\delta^\prime=\delta\left(\frac{s_x}{\Lambda}\right)^3, ~~
\alpha^\prime = \alpha \frac{d^{}}{\Lambda}~, ~~ \beta^\prime =
\beta \frac{d^{}}{\Lambda}~,~~
\gamma^\prime=\gamma\frac{s_z}{\Lambda}~.
\end{eqnarray}
As one can see, $M_\nu^{3\times 3}$ is
$\mu\hif\tau$ symmetric if $\varepsilon^{}_{d}=0(d_1 = d_2)$ holds.

Here, let us roughly estimate the magnitude of model parameters.
From the charged lepton sector, we obtain $s_{x,y} / \Lambda \simeq
(2\cdots 5)\times 10^{-3}$ for $Y_{e,x,y} = {\cal O}(1)$. Suppose
$s_x / \Lambda = 10^{-3}$ and $s_x \simeq s_y \simeq s_z \simeq d_1
\simeq d_2$ for simplicity, then $m_c={\cal O}(1)~\gev$ results in
$M_1={\cal O}(1)~\kev$ for the lightest sterile neutrino. Since
$M_{2,3}\simeq m_a$ and $m_a={\cal O}(1)~\gev$,
$(\alpha^{\prime},~\beta^{\prime},~\gamma^{\prime})$ need to be
${\cal O}(10^{-7.5})$ in order to reproduce realistic active
neutrino masses $m_\nu ={\cal O}(10^{-2})~\ev$, and they correspond
to $(\alpha,~\beta,~\gamma) = {\cal O}(10^{-4.5})$. If we assume the
same value for $\delta$ as well, then we gain $\delta^{\prime} =
{\cal O}(10^{-13.5})$.

In the approximation of $\delta^{\prime} v\ll M_1$, the mass matrix
$M_{\nu}^{4\times 4}$ in Eq. (\ref{eq:4by4}) can further be diagonalized by
a $4 \times 4$ neutrino mixing matrix parametrized as
\begin{eqnarray}
V_{\nu}\equiv V_{1}V_{0}\simeq\bmx{cc} i\sqrt{1-R^{}_{}R^{+}_{}} & R^{}_{} \\
-iR^{+}_{} & \sqrt{1-R^{+}_{}R^{}_{}} \emx~ \cdot
\bmx{cc} V_{\rm A} & 0 \\
0 & 1 \emx~\,,
\end{eqnarray}
where $R^{}_{}$ is a $3 \times 1$ mixing matrix which induces the
active-sterile mixing angles
\begin{eqnarray}
R^{}_{}\simeq \Delta/{M}^{}_{1}\equiv(\delta^{\prime}
v/{M}^{}_{1},0,0)^{T}\, , \label{eq:ASmixing}
\end{eqnarray}
and $V_{\rm A}$ is a unitary $3 \times 3$ mixing matrix
diagonalizing the mass matrix of three active neutrinos. We can use
$V_{1}$ to eliminate the $\Delta$ and $\Delta^{T}$ terms of
$M_{\nu}^{4\times 4}$,
\begin{eqnarray}
V^{\dag}_{1}M^{4 \times 4}_{\nu}V^{\ast}_{1}\simeq \bmx{cc} M^{3
\times 3}_{\nu}+R^{}_{}{M}^{}_{1}R^{T}_{}& 0 \\
0 & {M}^{}_{1} \emx~,
\end{eqnarray}
and $V_{A}$ can be determined by the relation $V^{\dag}_{\rm A}(M^{3
\times 3}_{\nu}+R^{}_{}{M}^{}_{1}R^{T}_{})V^{\ast}_{\rm A}={\rm
diag}\,\{\lambda^{}_{1},\lambda^{}_{2},\lambda^{}_{3}\}$. Because
the order of $R^{}_{}{M}^{}_{1}R^{T}_{}$ [i.e., $\delta^{\prime 2}
v^{2}/{M}^{}_{1}={\cal O}(10^{-8})~\ev$] is much smaller than that
of $M^{3 \times 3}_{\nu}$, we can safely neglect its effects in the
active neutrino part. The only exception is the contribution to
$\lambda^{}_{1}$ because it is vanishing when $M^{3 \times 3}_{\nu}$
is taken into account alone.

\section{Neutrino masses and mixing}
\subsection{ACTIVE NEUTRINO MASSES AND MIXING}
As we mentioned, if $\varepsilon_d =0$ holds, $M^{3 \times 3}_{\nu}$
is $\mu\hif\tau$ symmetric and it results in the vanishing
$\theta_{13}^{}$ and maximal $\theta_{23}^{}$ for active neutrinos.
In fact, these predictions are roughly compatible with the recent
neutrino oscillation data \cite{Fogli,GMS,STV}, which indicate a
small $\theta_{13}^{}$ and a nearly maximal $\theta_{23}^{}$. Thus,
$\varepsilon_{d}$ may be expected to be not so large, so that we
here treat $\varepsilon_{d}$ as a small parameter and employ the
perturbation calculations. The leading (i.e., $\mu\hif\tau$
symmetric) term can be diagonalized by
\begin{eqnarray}
V^{0}_{\rm A}
 &=&\bmx{ccc}
c_\theta\,e^{i\rho} & s_\theta\,e^{i\rho} & 0\\
 -\frac{s_\theta}{\sqrt{2}}\,e^{i\sigma} & \frac{c_\theta}{\sqrt{2}}\,e^{i\sigma} & -\frac{1}{\sqrt{2}}\\
 -\frac{s_\theta}{\sqrt{2}}\,e^{i\sigma} & \frac{c_\theta}{\sqrt{2}}\,e^{i\sigma} & \frac{1}{\sqrt{2}}
 \emx\,,
 \label{eq:V0}
\end{eqnarray}
where $c_\theta\equiv\cos\theta\,$ and $s_\theta\equiv\sin\theta$, and the mixing parameters are given by
\begin{eqnarray}
\rho = {\rm arg}{\alpha^{\prime}},~~ \sigma = {\rm
arg}(\beta^{\prime}+\gamma^{\prime}),~~
\sin{\theta}=\frac{\sqrt{2}|\alpha^{\prime}|}{\sqrt{|\beta^{\prime}+\gamma^{\prime}|^2+2|\alpha^{\prime}|^2}}\;,
\end{eqnarray}
together with the eigenvalues
\begin{eqnarray}
\lambda^{0}_{1} = 0,~~ \lambda^{0}_{2} =
\frac{|\beta^{\prime}+\gamma^{\prime}|^2+2|\alpha^{\prime}|^2}{m_{a}}v^2,~~
\lambda^{0}_{3}=-\frac{(\beta^{\prime}-\gamma^{\prime})^2}{m_{a}}v^2~~.
 \label{eq:mass}
\end{eqnarray}
Notice that $\rho$ and $\sigma$ are unphysical phases and do not
affect any observables, though they appear in the following
expressions. After including the correction term and doing
perturbative calculations, we get the neutrino mixing angles in the
standard parametrization as
\begin{eqnarray}
\sin\theta^{}_{12}&\simeq& \sin\theta^{}_{}+{\cal
O}(\varepsilon^{2}_{d})\nonumber \\
\tan\theta^{}_{23}&\simeq&
\left|1+4\beta^{\prime}\gamma^{\prime}\frac{e^{2i\sigma}}{\lambda^{0}_{3}-\lambda^{0}_{2}}
\frac{\varepsilon_{d}v^{2}}{m_{a}}\right|+{\cal
O}(\varepsilon^{2}_{d})
\simeq \left|1-\frac{4\beta^{\prime}\gamma^{\prime}e^{2i\sigma}}{(\beta^{\prime}-\gamma^{\prime})^2}
\varepsilon_{d}\right|
\nonumber\\
\sin\theta^{}_{13}&\simeq&\left|\sqrt{2}\alpha^{\prime}(2\beta^{\prime}-\gamma^{\prime})
\frac{e^{2i\rho}}{\lambda^{0}_{3}-\lambda^{0}_{2}}\frac{\varepsilon_{d}v^{2}}{m_{a}}\right|+{\cal
O}(\varepsilon^{2}_{d})
\simeq \left|\frac{\sqrt{2}\alpha^{\prime}(2\beta^{\prime}-\gamma^{\prime})}
{(\beta^{\prime}-\gamma^{\prime})^2}\varepsilon_{d}\right|\,,
\label{eq:ana-mix}
\end{eqnarray}
and the Jarlskog invariant parameter $J$ as
\begin{equation}
J\simeq\frac{\sqrt{2}}{2}\frac{|\alpha^{\prime}(2\beta^{\prime}-\gamma^{\prime})|}
{|\beta^{\prime}-\gamma^{\prime}|^2}\varepsilon_{d}\sin\phi+{\cal
O}(\varepsilon^{2}_{d})\,\,,
\end{equation}
where $\phi={\rm
arg}(2\alpha^{\prime}+\beta^{\prime}+2\gamma^{\prime})$. From Eq.
(\ref{eq:ana-mix}), one can check the recovery of the $\mu\hif\tau$
symmetry ($\theta_{13}^{}=0^\circ$ and $\theta_{23}^{}=45^\circ$) in
the limit of $\varepsilon_d = 0$. The corrections to the three
eigenvalues are vanishing in the order of ${\cal
O}(\varepsilon^{}_{d})$, so the active neutrino masses are obtained
by taking absolute values for Eq. (\ref{eq:mass}), and this model
predicts a normal mass hierarchy for three active neutrinos. Note
that the vanishing of $\lambda_{1}$ should be kept in all orders of
perturbations and is a generic property of the minimal seesaw models
\cite{Mini}. A nonvanishing mass of $\nu_1$ can be generated from
the lightest sterile neutrino contribution:
\begin{equation}
\lambda_{1}\simeq\delta^{\prime 2} v^{2}/{M}^{}_{1},
\end{equation}
but the corresponding effects are negligibly small and can be safely
ignored for all the other mass and mixing parameters.

\subsection{NUMERICAL ANALYSIS}
Instead of perturbative calculations, we here numerically
diagonalize Eq. (\ref{eq:3by3}) and compute $\theta_{13}$,
$\theta_{23}$ and $J$. From the recent global analysis \cite{STV} of
the neutrino oscillation data, in our calculations, we refer to the
following best-fit values and $3\sigma$ error bounds:
\begin{eqnarray}
&&\Delta m_{21}^2 =
\left( 7.59^{+0.60}_{-0.50} \right)
\times 10^{-5}_{}~~\ev^2,
~~~
\Delta m_{31}^2 =
\left( 2.50^{+0.26}_{-0.36} \right)
\times 10^{-3}_{}~~\ev^2, \nonumber \\
&&\sin^2\theta_{12}^{} =
0.312^{+0.048}_{-0.042},
~~~
\sin^2\theta_{23}^{} =
0.52^{+0.12}_{-0.13},
~~~
\sin^2\theta_{13}^{} =
0.013^{+0.022}_{-0.012},
\label{eq:gfit}
\end{eqnarray}
for the normal neutrino mass hierarchy.
\begin{figure}[ht]
\begin{center}
\includegraphics*[width=0.49\textwidth]{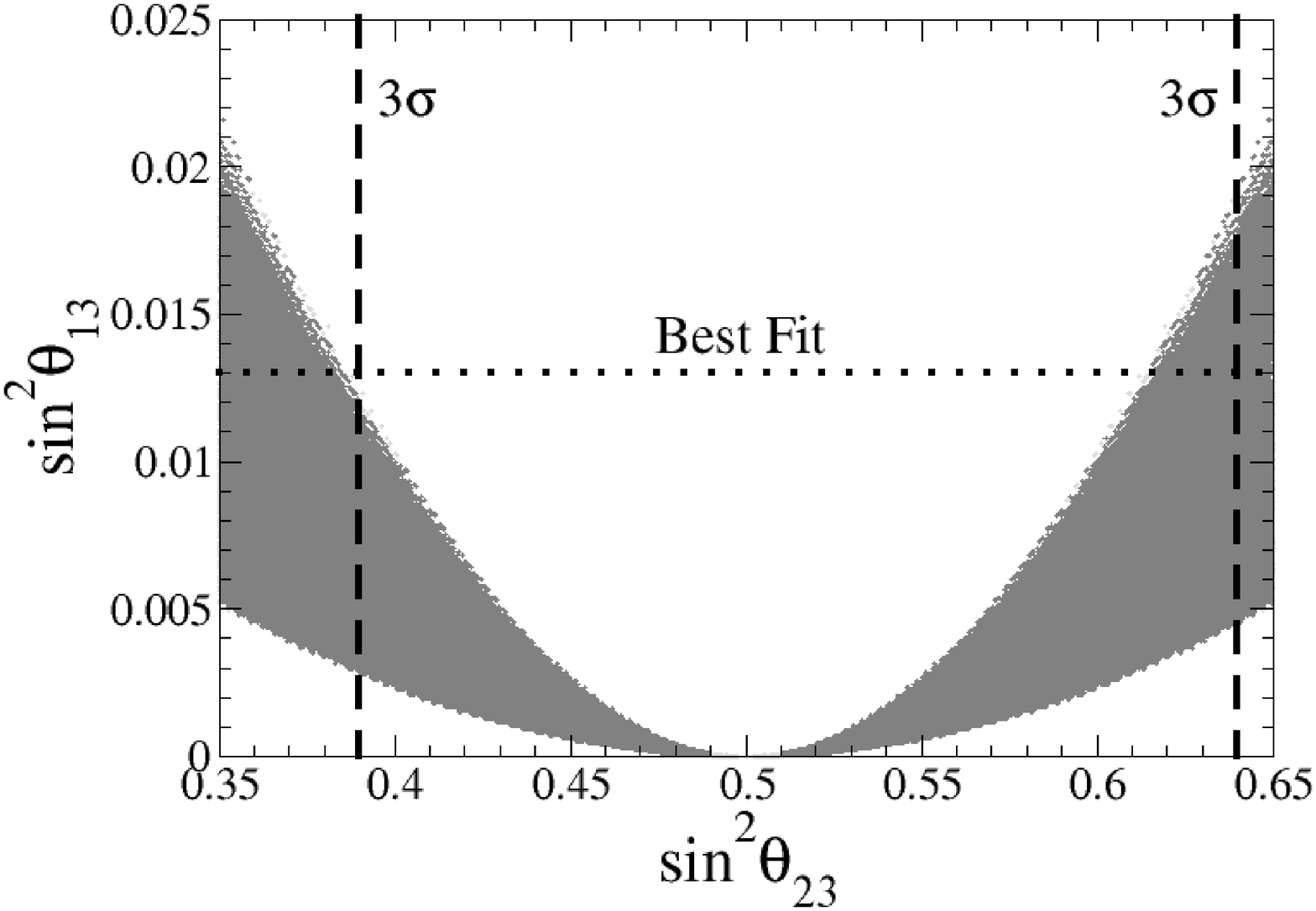}
\includegraphics*[width=0.49\textwidth]{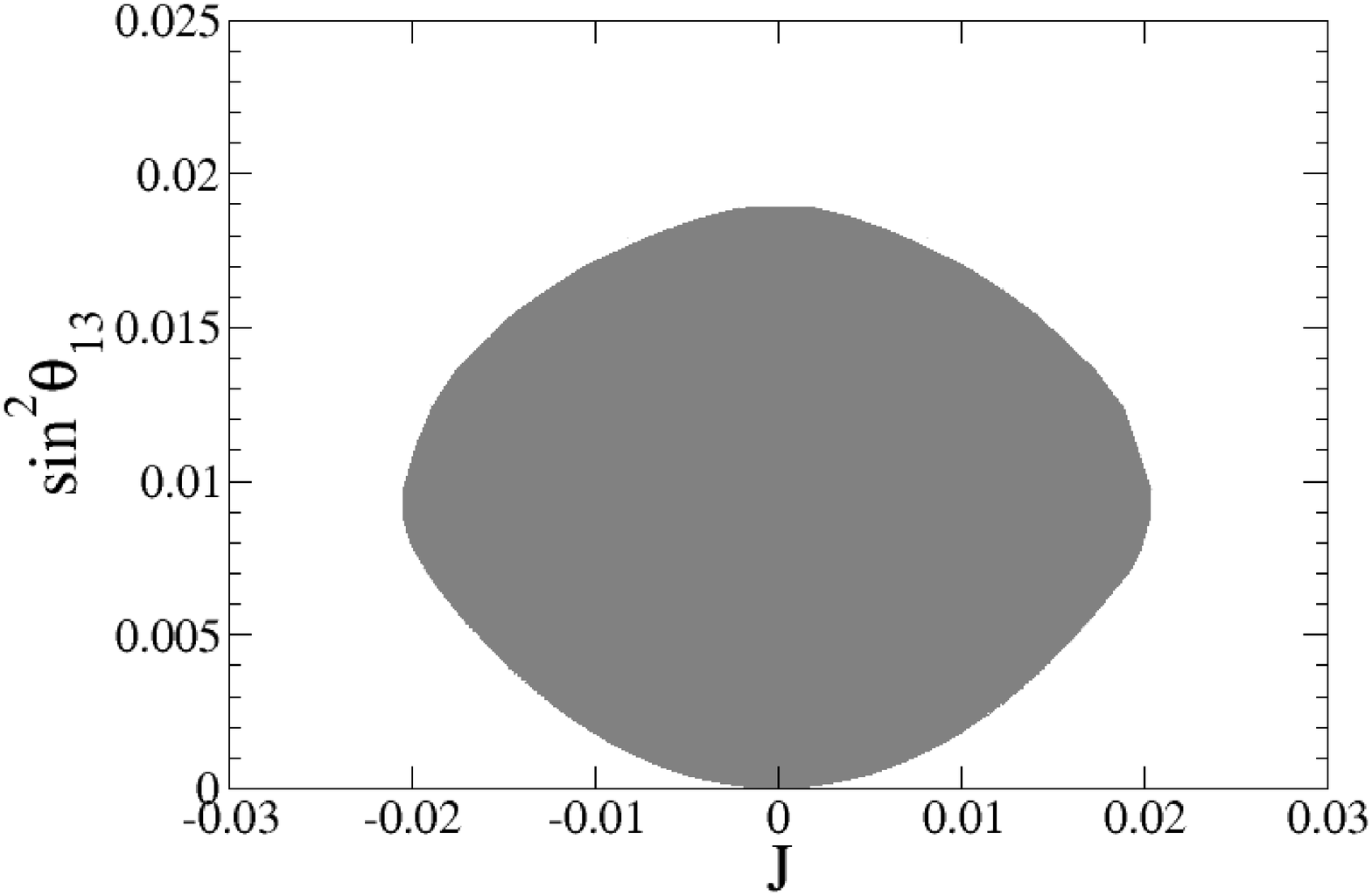}
\vspace{-0.3cm} \caption{\footnotesize $\sin^2\theta_{13}^{}$ vs
$\sin^2\theta_{23}^{}$ (left panel) and $\sin^2\theta_{13}^{}$ vs
$J$ (right panel), where the horizontal dotted line displays the
best-fit value of $\sin^2\theta_{13}^{}$ while the vertical dashed
lines express the $3\sigma$ upper and lower bounds of
$\sin^2\theta_{23}$ from Eq. (\ref{eq:gfit}). } \label{fig1}
\end{center}
\end{figure}
In Figure \ref{fig1}, we plot $\sin^2\theta_{13}^{}$ as functions of
$\sin^2\theta_{23}^{}$ (left panel) and $J$ (right panel) with the
$3\sigma$ constraints of $\Delta m_{21}^2$, $\Delta m_{31}^2$ and
$\sin^2\theta_{12}^{}$. Besides, the $3\sigma$ bound of
$\sin^2\theta_{23}^{}$ is also imposed in the
$\sin^2\theta_{13}^{}-J$ plane (right panel). As one can see from
the figure, the predicted regions can be within the $3\sigma$
ranges, and $\sin^2\theta_{13}^{}$ can deviate from $0$, which is
favored by the recent T2K \cite{T2K} and MINOS \cite{MINOS} results.
However, a large $\sin^2\theta_{13}^{}$ is always accompanied with a
large deviation of $\sin^2\theta_{23}^{}$ from $0.5$. For instance,
the best-fit value of $\sin^2\theta_{13}^{}$ can be accounted for at
around $|\sin^2\theta_{23}^{}-0.5|\simeq0.11$, but it is almost the
edge of the $3\sigma$ bound\footnote{We refer to Ref. \cite{13-23}
for the realizations of a large $\theta_{13}^{}$ together with a
small deviation of $\theta_{23}^{}$ from the maximality. }. Notice
that we have checked that the $\mu\hif\tau$ symmetry breaking
parameter $\varepsilon_d^{}$, which is defined below Eq.
(\ref{eq:3by3}), is at most $0.4$ and that the analytical
expressions in Eqs. (\ref{eq:mass}) and
(\ref{eq:ana-mix}) approximately agree with the numerical results in
Figure \ref{fig1}.

\subsection{ACTIVE-STERILE MIXING}
From Eq. (\ref{eq:ASmixing}) and the parameter estimates in section
III, the active-sterile mixing angles $\Theta^{}_{i}$ defined by the
elements of the $R^{}_{}$ matrix (i.e.,
$\Theta^{}_{i}\simeq|R^{}_{i1}|$) can be derived as
\begin{eqnarray}
\Theta^{2}_{1}
\simeq\left|\frac{\lambda_{1}}{{M}^{}_{1}}\right|\simeq10^{-11}~~,
\end{eqnarray}
and $\Theta^{2}_{2}\simeq\Theta^{2}_{3}\simeq 0$, which are well
below the upper bounds from astrophysical and cosmological
observations \cite{Review}. Furthermore, it is also consistent with
the requirement of correct DM abundance for the mechanism of
resonant active-sterile oscillations with nonzero lepton asymmetries
\cite{Shi}.
To achieve the right DM abundance with the nonresonant mechanism
\cite{DW}, we need $\delta$ to be one order of magnitude larger than
$(\alpha,~\beta,~\gamma)$ in order to achieve $\Theta^{2}_{1}\simeq
10^{-9}$ \cite{Review}. These tiny active-sterile mixing angles make
the detection of the WDM particle rather dim and remote with both
the X-ray observations \cite{Xray} and the captures on beta-decaying
or electron-capture-decaying nuclei \cite{Liao,LX11}.

\section{Conclusions}
In this work, we have proposed a $Q_6$ flavor symmetry realization
for the $\nu$MSM in the presence of two auxiliary $Z_N$ symmetries
and succeeded in naturally explaining the lightness of the $\kev$
sterile neutrino and the mass degeneracy of the two heavier sterile
neutrinos. A normal hierarchical mass spectrum and an approximately
$\mu$-$\tau$ symmetric mass matrix are predicted for three active
neutrinos. Nonzero $\theta_{13}$ can be obtained together with a
deviation of $\theta_{23}$ from the maximality, where both mixing
angles are consistent with the latest global data including T2K and
MINOS results. Finally, we have derived a tiny active-sterile mixing
related to the mass ratio between the lightest active and lightest
sterile neutrinos.

The $\nu$MSM can explain the active neutrino masses, the candidate
of dark matter and the baryon asymmetry in the Universe in a unified
and elegant way. Although there are already some models in which the
$\nu$MSM is extended by flavor symmetries, our model has more direct
connections with the masses and mixing patterns of three active
neutrinos. Our realization can also be modified to accommodate the
$\ev$ scale sterile neutrinos \cite{eVmodel}, which are more or less
hinted at by current experimental \cite{Schwetz4} and cosmological
\cite{Raffelt} data. We shall examine this case with a specific
flavor model elsewhere \cite{TA}.

\section*{ACKNOWLEDGMENTS}
This work was supported in part by the National Natural Science
Foundation of China under Grant No. 11135009, by the Chinese Academy
of Sciences Fellowship for Young International Scientists (T.A.) and
by the China Postdoctoral Science Foundation under Grant No.
20100480025 (Y.F.L.).

\appendix\section{Basics of $Q_6$}
$Q_6$ consists of four singlet and two doublet irreducible
representations,
\begin{eqnarray}
\true,~ \ip,~ \ipp,~ \ippp,~ \two,~ \tp,
\end{eqnarray}
and twelve elements,
\begin{eqnarray}
E, ~R_6^{}, ~R_6^2, \cdots R_6^5, ~P_Q^{}, ~ R_6^{}P_Q^{},
~R_6^2P_Q^{}, \cdots R_6^5P_Q^{},
\end{eqnarray}
where $E$ stands for the unit matrix.
The representation matrices
of $R_6^{}$ and $P_Q^{}$ for each representation are give by
\begin{eqnarray}
\begin{array}{lllll}
\true & \ddag & R_6^{}= 1 & & P_Q^{}= 1 \\
\ip   & \ddag & R_6^{}= 1 & & P_Q^{}=-1 \\
\ipp  & \ddag & R_6^{}=-1 & & P_Q^{}=-i \\
\ippp & \ddag & R_6^{}=-1 & & P_Q^{}= i \\
\two  & \ddag & R_6^{}= \bmx{cc}
\omega_6^{} & 0 \\
0 & \omega_6^{-1} \emx  & & P_Q^{}= \bmx{cc}
0 & i \\
i & 0
\emx \\
\tp & \ddag & R_6^{}= \bmx{cc}
\omega_6^2 & 0 \\
0 & \omega_6^{-2} \emx  & & P_Q^{}= \bmx{cc}
0 & 1 \\
1 & 0 \emx
\end{array}
\end{eqnarray}
with $\omega_6^{}=\exp{\left\{i\frac{2\pi}{6}\right\}}$. Note that
$\true$, $\ip$ and $\tp$ are real representations, $\two$ is a
pseudoreal representation and $\ippp=({\ipp})^*$ are complex
representations.
\begin{table}
\begin{tabular}{|l||c|c|c|c|c|}\hline
        & $\ip$   & $\ipp$  &
$\ippp$ & $\two$  & $\tp$ \\ \hline\hline
$\ip$   & $\true$ & $\ippp$ & $\ipp$  & $\two$  & $\tp$ \\ \hline
$\ipp$  & ~*~     & $\ip$   & $\true$ & $\tp$   & $\two$ \\ \hline
$\ippp$ & ~*~     & ~*~     & $\ip$   & $\tp$   & $\two$ \\ \hline
$\two$  & ~*~     & ~*~     & ~*~     & $\true \oplus \ip \oplus
\tp$ & $\ipp \oplus \ippp \oplus \two$ \\ \hline $\tp$   & ~*~     &
~*~     & ~*~     & ~*~     & $\true \oplus \ip \oplus \tp$ \\
\hline
\end{tabular}
\caption{A table of the tensor products of $Q_6$.}
\end{table}
The tensor products among the irreducible representations are
summarized in Table II. Especially, the products of two doublets are
defined as follows.
\begin{eqnarray}
\begin{array}{ccccccccc}
\bmx{c} x_1 \\ x_2 \emx & \otimes & \bmx{c} y_1 \\ y_2 \emx & = &
(x_1y_2-x_2y_1) & \oplus & (x_1y_2+x_2y_1) & \oplus &
\bmx{c} x_1y_1 \\ -x_2y_2 \emx \\
\two & \otimes & \two & = & \true & \oplus &
\ip & \oplus & \tp \\
&&&&&&&& \\
\bmx{c} x_1 \\ x_2 \emx & \otimes & \bmx{c} y_1 \\ y_2 \emx & = &
(x_1y_2+x_2y_1) & \oplus & (x_1y_2-x_2y_1) & \oplus &
\bmx{c} x_2y_2 \\ x_1y_1 \emx \\
\tp & \otimes & \tp & = & \true & \oplus &
\ip & \oplus & \tp\\
&&&&&&&& \\
\bmx{c} x_1 \\ x_2 \emx & \otimes & \bmx{c} y_1 \\ y_2 \emx & = &
(x_1y_1-x_2y_2) & \oplus & (x_1y_1+x_2y_2) & \oplus &
\bmx{c} x_2y_1 \\ x_1y_2 \emx \\
\two & \otimes & \tp & = & \ipp & \oplus & \ippp & \oplus & \two
\end{array}
\end{eqnarray}

\end{document}